\documentstyle{mn}

\newif\ifAMStwofonts

\ifoldfss
  \ifCUPmtlplainloaded \else
    \NewTextAlphabet{textbfit} {cmbxti10} {}
    \NewTextAlphabet{textbfss} {cmssbx10} {}
    \NewMathAlphabet{mathbfit} {cmbxti10} {} 
    \NewMathAlphabet{mathbfss} {cmssbx10} {} 
  \fi
  \ifAMStwofonts
    \ifCUPmtlplainloaded \else
      \NewSymbolFont{upmath} {eurm10}
      \NewSymbolFont{AMSa} {msam10}
      \NewMathSymbol{\upi}     {0}{upmath}{19}
      \NewMathSymbol{\umu}     {0}{upmath}{16}
      \NewMathSymbol{\upartial}{0}{upmath}{40}
      \NewMathSymbol{\leqslant}{3}{AMSa}{36}
      \NewMathSymbol{\geqslant}{3}{AMSa}{3E}

    \fi
  \fi
\fi 

\ifnfssone
  \newmathalphabet{\mathit}
  \addtoversion{normal}{\mathit}{cmr}{m}{it}
  \addtoversion{bold}{\mathit}{cmr}{bx}{it}
  \newmathalphabet{\mathbfit} 
  \addtoversion{normal}{\mathbfit}{cmr}{bx}{it}
  \addtoversion{bold}{\mathbfit}{cmr}{bx}{it}
  \newmathalphabet{\mathbfss} 
  \addtoversion{normal}{\mathbfss}{cmss}{bx}{n}
  \addtoversion{bold}{\mathbfss}{cmss}{bx}{n}
  \ifAMStwofonts
    \ifCUPmtlplainloaded \else
      \UseAMStwoboldmath
      \makeatletter
      \new@mathgroup\upmath@group
      \define@mathgroup\mv@normal\upmath@group{eur}{m}{n}
      \define@mathgroup\mv@bold\upmath@group{eur}{b}{n}
      \edef\UPM{\hexnumber\upmath@group}
      \new@mathgroup\amsa@group
      \define@mathgroup\mv@normal\amsa@group{msa}{m}{n}
      \define@mathgroup\mv@bold\amsa@group{msa}{m}{n}
      \edef\AMSa{\hexnumber\amsa@group}
      \makeatother
      \mathchardef\upi="0\UPM19
      \mathchardef\umu="0\UPM16
      \mathchardef\upartial="0\UPM40
      \mathchardef\leqslant="3\AMSa36
      \mathchardef\geqslant="3\AMSa3E
    \fi
  \fi
\fi 
\ifnfsstwo
  \DeclareMathAlphabet{\mathbfit}{OT1}{cmr}{bx}{it}
  \SetMathAlphabet\mathbfit{bold}{OT1}{cmr}{bx}{it}
  \DeclareMathAlphabet{\mathbfss}{OT1}{cmss}{bx}{n}
  \SetMathAlphabet\mathbfss{bold}{OT1}{cmss}{bx}{n}
  \ifAMStwofonts
    \ifCUPmtlplainloaded \else
      \DeclareSymbolFont{UPM}{U}{eur}{m}{n}
      \SetSymbolFont{UPM}{bold}{U}{eur}{b}{n}
      \DeclareSymbolFont{AMSa}{U}{msa}{m}{n}
      \DeclareMathSymbol{\upi}{0}{UPM}{"19}
      \DeclareMathSymbol{\umu}{0}{UPM}{"16}
      \DeclareMathSymbol{\upartial}{0}{UPM}{"40}
      \DeclareMathSymbol{\leqslant}{3}{AMSa}{"36}
      \DeclareMathSymbol{\geqslant}{3}{AMSa}{"3E}
    \fi
  \fi
\fi 

\ifCUPmtlplainloaded \else
  \ifAMStwofonts \else 
    \def\upi{\pi}
    \def\umu{\mu}
    \def\upartial{\partial}
  \fi
\fi

\title[GX\,17+2]
  {GX\,17+2: X-ray spectral and timing behaviour of a bursting Z source}

\author[Kuulkers et al.]
  {E.~Kuulkers,$^1$\thanks{{\it Present address:\/} 
  Astrophysics, University of Oxford, Nuclear and Astrophysics Laboratory, 
  Keble Road, Oxford OX1 3RH, United Kingdom} 
  M.~van der Klis,$^1$ T.~Oosterbroek,$^1$
  J.~van Paradijs$^{1,2}$ and 
  \newauthor
  W.~H.~G.~Lewin$^3$\\
  $^1$Astronomical Institute ``Anton Pannekoek'', University of Amsterdam
  and \\
  Center for High Energy Astrophysics,
  Kruislaan 403, 1098 SJ Amsterdam, The Netherlands\\
  $^2$Physics Department, University of Alabama, Huntsville, AL 35899, U.S.A.\\
  $^3$Massachusetts Institute of Technology, 37-627, Cambridge, MA~02139, U.S.A.}

\date{Accepted. Received}
\pagerange{\pageref{firstpage}--\pageref{lastpage}}
\pubyear{1996}

\begin{document}

\label{firstpage}

\maketitle

\begin{abstract}

We investigated the properties of the Z pattern of the low-mass X-ray
binary GX\,17+2 in the X-ray colour-colour diagrams (CDs) and 
hardness-intensity diagrams (HIDs), and of the power spectra as
a function of position in the Z. We used all {\it EXOSAT} ME data on GX\,17+2,
a total of 6 observations during 1983-1986. 
We find that the Z pattern in the CD does not 
move to within $\sim$2 per cent between observations separated by 
$\sim$1 day to $\sim$1.5 years.
The power spectra and the characteristics of the Z pattern both support 
the division of the Z sources into two sub-groups, GX\,17+2 being in the 
group together with Sco\,X-1 and GX\,349+2.

It has been noted previously that the maximum frequency of the horizontal 
branch quasi-periodic oscillations in GX\,17+2 is lower than for Cyg\,X-2, 
GX\,5--1 and GX\,340+0. We suggest that an asymmetry in the magnetic field  
giving rise to polar caps with different emission characteristics 
may be the origin of this. This may also explain the occurrence of bursts in 
GX\,17+2.

We also report the detection of two new bursts in the {\it EXOSAT} data of
GX\,17+2, which we interpret as type~I bursts.
No regular pulsations are observed during these two bursts nor during the 
bursts previously detected in GX\,17+2 with {\it EXOSAT}, with upper limits to the 
pulsed amplitude of $\sim$1.5 per cent for pulsations with a frequency below 512\,Hz.
Three of the four bursts occurred in the normal branch, while one occurred 
in the lower part of the flaring branch.

The Poisson level in power spectra of data processed
by the on-board computer of {\it EXOSAT} is found to deviate from that expected from
simple variable dead time effects in GX\,17+2 and also in several 
other sources. We present corrections
to the usually employed relations between count rates observed with 
{\it EXOSAT} and the Poisson level.

\end{abstract}


\section{Introduction}

Studies of the broad-band X-ray spectral and correlated timing behaviour 
of the brightest persistent low-mass
X-ray binaries (LMXBs) have shown that GX\,17+2 is one of the six known
Z sources (Hasinger \&\ van der Klis 1989), which share many 
characteristics. In the X-ray colour-colour
diagram (CD) and hardness-intensity diagram (HID) the sources trace out a 
roughly ``Z'' shaped track. The limbs of the Z are called horizontal branch 
(HB), normal branch (NB) and flaring branch (FB), 
from the top limb to the bottom limb, respectively. The sources move 
smoothly through the different branches, and do not jump from branch to 
branch. This one-dimensional motion through the Z is thought to be 
governed by the mass-accretion rate, $\dot{M}$, which increases from the HB, 
through the NB, to the FB (e.g.\ Hasinger et al.\ 1990). The Z sources are 
believed to accrete mass at a rate near the Eddington limit (e.g.\ Lamb 1989).

In each branch, a Z source shows a characteristic power spectrum 
(Hasinger \&\ van der Klis 1989). In the HB one sees intensity
correlated high-frequency 
(15--55\,Hz) quasi-periodic oscillations (QPO), called horizontal-branch 
oscillations (HBO), together with associated band-limited noise, 
called low-frequency noise (LFN), which dominates the region from 
$\sim$1\,Hz to the HBO frequency. In the 
NB one sees another type of QPO at a lower frequency (4--7\,Hz), called normal-branch oscillations
(NBO). NBO have been seen to occur 
simultaneously with HBO (e.g.\ Hasinger et al.\ 1990, Lewin et al.\ 1992). 
In some Z sources (\mbox{Sco\,X-1}, see e.g.\ 
Dieters \&\ van der Klis [1996], and 
GX\,17+2, see e.g.\ Penninx et al.\ [1990]), the NBO merge smoothly into higher 
frequency (up to $\sim$20\,Hz) QPO, called flaring branch oscillations (FBO) 
when the source moves onto the FB. Recently, kHz QPO have been discovered
in the Z source Sco\,X-1 (van der Klis et al.\ 1996), whose properties 
also seem to depend on position in the Z.

Two noise components have been seen to occur in all three branches.
One is the very-low-frequency noise (VLFN), dominating the region below 
$\sim$1\,Hz. VLFN generally increases towards the FB.
The other component is the high-frequency noise (HFN), which dominates the 
region above $\sim$20\,Hz. HFN generally decreases towards the FB (see 
Dieters \&\ van der Klis 1996, and references therein).

It has become clear that on the basis of their spectral and timing behaviour 
the Z sources can be divided into two categories
(Hasinger \&\ van der Klis 1989, Penninx et al.\ 1991, Kuulkers et al.\ 1994a), 
the difference between them may be due to a difference in orbital 
inclination angle 
(Hasinger \&\ van der Klis 1989, Hasinger, Priedhorsky \&\ Middleditch
1989, Hasinger et al.\ 1990, Kuulkers et al.\ 1994a, Kuulkers 1995, 
Kuulkers \&\ van der Klis 1995, 1996, Kuulkers, van der Klis \&\ Vaughan 1996)
or magnetic field strength of the neutron star 
(Psaltis, Lamb \&\ Miller 1995)\footnote{If it is due to orbital 
inclination the Z sources
would be intrinsically the same, but appear differently; if it is due to
magnetic field, then they would be intrinsically different.}.
The Z sources Cyg\,X-2, GX\,5--1 and GX\,340+0 form one group, which we shall 
call the Cyg-like sources, while Sco\,X-1, GX\,17+2 and GX\,349+2 form 
another group, which we shall call the Sco-like sources.
The Cyg-like sources have been proposed to have either a higher inclination,
or a higher magnetic field, than the Sco-like sources.

The only direct evidence that the compact stars in Z sources are neutron 
stars is the observation of X-ray bursts from GX\,17+2 (Tawara et al.\ 1984, 
Kahn \&\ Grindlay 1984, Sztajno et al.\ 1986) and Cyg\,X-2 (Kahn \&\ Grindlay 
1984, Kuulkers, van der Klis \&\ Van Paradijs 1995, Wijnands et al.\ 1997, 
Smale et al.\ 1996).
The irregular occurrence of these X-ray bursts, and the lack of bursts in 
other Z sources has been attributed to the high mass accretion rate $\dot{M}$ 
in these systems (see the review by Lewin, Van Paradijs \&\ Taam 1993).

Recently, the {\it EXOSAT} data of three Z sources have been analyzed in a
detailed and homogeneous way (GX\,5--1: Kuulkers et al.\ 1994a, GX\,340+0: 
Kuulkers \&\ van der Klis 1996, Sco\,X-1: Dieters \&\ van der Klis 1996). 
The present work reports the results of a similar comprehensive study of the 
broad-band X-ray spectral and timing behaviour of GX\,17+2. Parts of the 
{\it EXOSAT} data on this source have already been reported by Langmeier et al.\ 
(1986), Langmeier, Hasinger \&\ Tr\"umper (1990), Sztajno et al.\ (1986), 
Stella, Parmar \&\ White (1987), Schulz, Hasinger \&\ Tr\"umper (1989) and 
Hasinger \&\ van der Klis (1989). 
In our study we discuss the overall broad-band spectral and timing 
behaviour within the framework of the magnetospheric model for Z sources and
of the idea that GX\,17+2 is a Z source with a relatively low inclination.
We also report previously unseen bursts, and place all {\it EXOSAT} bursts within 
the Z-diagram (a preliminary report of these bursts was given by Kuulkers 
et al.\ 1994b).

\section{Observations}

We analysed all available {\it EXOSAT} medium energy experiment (ME) argon data 
(Turner, Smith \&\ Zimmerman 1981, White \&\ Peacock 1988) of GX\,17+2. 
In Table~1 we present the observation log of GX\,17+2.
The ME had eight detectors, each of which consisted of two layers, 
an argon filled chamber in front of a xenon 
filled chamber\footnote{Early on 1985 day 232 
one of the eight detectors ceased to operate.}. Data 
were obtained in different observation modes, 
with good spectral and low time resolution (HER2, HER3, HER4, HER5; the 
different HER modes gave different options to transmit data from individual 
detectors or combined from more detectors), or with 
high time and no spectral resolution (HTR3, HTR4, HTR5). Alternatively, one 
could obtain data in two (e.g.\ 1985 day 258/259) or four energy bands 
(e.g.\ 1986 day 093/094) at a high time resolution (HER7).
A HER mode could be run simultaneously with one of the 
HTR modes. Either one half of the array of eight detectors (``half 1'' [H1] 
or ``half 2'' [H2]) was pointed at the source 
while the other half monitored the background, or the whole array (WA) was 
pointed at the source (see Table 1).

The HER modes were processed by the {\it EXOSAT} onboard computer (OBC), which 
produced large dead-time effects. In our analysis of CDs 
and HIDs (see below) we determine the dead-time factor from the qualified 
event rates (collected each 32s for each detector), which
are (almost) unaffected by dead time. Of the HTR modes, only the HTR4 mode 
is processed by the OBC. The HTR4 mode provides the count rate in only one 
(chosen) energy band at the 
highest time resolution available with {\it EXOSAT} ($\sim$0.25\,ms). The 
HTR4 mode observations on 1985 day 258/259 and on 1986 day 093/094 
provided the count rate in the 1.4--11.4\,keV band and in the 
1.2--19.9\,keV band, respectively.

On 1986 February 13th (Thursday, day 044) problems due to atmospheric 
drag caused by orbital decay started with the {\it EXOSAT} attitude 
and orbit control system (AOCS), see White (1986). Between March 27th 
(day 086) and April 9th (day 099) stable pointing was regained within 
5\,arcmin. This resulted in a varying collimator response.
In principle one could use the star tracker attitude control data
(collected every 8 or 16\,s) to correct for this. However, due to problems 
during the 1986 day 093/094 observations with the star trackers and gyros
this was not possible for this observation period. We therefore used these 
data only to produce a CD and not for the HIDs. After April 9th (day 099) 
stable pointing was lost again, which resulted in a complete loss of the 
satellite. On May 9th (day 129) {\it EXOSAT} re-entered the Earth's atmosphere 
(White 1986).

\section{Analysis}

For the spectral analysis we use CDs and HIDs. These diagrams are
created using either three or four energy bands (e.g.\ Hasinger \&\ van der 
Klis 1989). The observations during 1986 day 093/094 provided four 
energy bands, so we decided to analyse all the data using 
energy boundaries corresponding to these bands. We found that three energy 
bands (i.e.\ combining two of the four bands) gave the best separation of the 
branches of GX\,17+2 in the CD: 1.2--4.7\,keV, 4.7--6.6\,keV and 
6.6--19.9\,keV. In the CD and HIDs the soft colour is defined as the ratio of 
the count rate in the second energy band to the count rate in the first 
energy band, while the hard colour is defined as the ratio of the count rate 
in the third energy band to that in the second energy band. The `intensity' 
used in the HIDs is the count rate per cm$^{2}$ in the 1.2--19.9\,keV band, 
corrected for dead 
time and collimator response (as redetermined by Kuulkers et al.\ 1994a).
The count rates were corrected for background.
All points in the CD and HIDs are 200\,s averages. 

In order to investigate the fast timing behaviour of GX\,17+2 we performed 
fast Fourier transforms (FFTs) on successive 16\,s and 512\,s blocks 
of data of all the high-time resolution argon data. 
Power spectra of blocks of 512\,s of data were used to study primarily the 
low-frequency behaviour $<$1\,Hz. In order to obtain a more accurate binning 
in position in the Z, we used the power spectra of 16\,s blocks of data. The 
latter power spectra were used to study primarily the high frequency ($>$1\,Hz) 
behaviour. The power spectra were grouped according to position in the Z and 
then averaged.

When analysing power spectra of Z sources, one 
encounters several power spectral components
(Hasinger \&\ van der Klis 1989):
a constant level due to counting noise (the so-called Poisson level) modified 
by dead-time processes, three ``noise'' components, VLFN, LFN and HFN,
and two kinds of QPO, HBO and NBO/FBO.
In some cases the HBO show a second harmonic, which is also 
represented by a Lorentzian. For completeness we give the functional shapes 
of these components (see also Hasinger \&\ van der Klis 1989, 
Kuulkers et al.\ 1994a):
\begin{description}
\item The VLFN is given as:
$P_{\rm VLFN}(\nu) = A_{\rm V}\nu^{-\alpha_{\rm V}}$, 
where $\nu$ is the frequency, $\alpha_{\rm V}$ the power-law index, and
$A_{\rm V}$ the normalization constant. 
\item The LFN and HFN are given as:
$P_{\rm Noise}(\nu) = A_{\rm N}\nu^{-\alpha_{\rm N}}e^{-\nu/\nu_{\rm N}}$,
where $\alpha_{\rm N}$ is the power-law index, $\nu_{\rm N}$ the cut-off 
frequency, and $A_{\rm N}$ the
normalization constant. In the Z sources $\alpha_{\rm N}$ was usually 
found to be consistent with zero for the HFN component (Hasinger \&\ van der 
Klis 1989, see also Dieters \&\ van der Klis 1996), and we fixed the 
parameter at this value. In Tables 2, 3a and 3b we use the subscript 
L and H for the LFN and HFN components, respectively.
\item The HBO, their harmonic, the NBO and the FBO are broad peaks with centroid frequencies
$\nu_{\rm c}$. These peaks are described by Lorentzians: 
$P_{\rm QPO}(\nu) = A_{\rm Q} [(\nu - \nu_{\rm Q})^2 + (\Delta\nu_{\rm Q}/2)^2]^{-1}$,
where $\Delta\nu_{\rm Q}$ is the full width at half maximum (FWHM) of the 
QPO and $A_{\rm Q}$ a normalization constant. 
\end{description}
The strengths of the various noise components are expressed in terms of the 
fractional rms amplitudes of the corresponding fluctuations in the time
series. They are calculated by integrating their contributions to the power
spectra over certain frequency ranges, as determined from fits of the 
functional shapes given above to the power spectra.
The VLFN was integrated in the 512\,s power spectra over the range
0.001 to 1\,Hz, and in the 16\,s power spectra over the range 0.01 to 1\,Hz. 
For the LFN and HFN we integrated over the range 0.01 to 100\,Hz.
Note that in the 16\,s power spectra we extrapolated the power spectral 
shapes from $\sim$0.06\,Hz down to 0.01\,Hz. 
The HER7 and HTR4 data suffer 
from large variable dead time effects. We corrected the 
fractional rms amplitudes of the power spectral components for these effects 
(see van der Klis 1989, Kuulkers et al.\ 1994a).

The Poisson levels of the power spectra from non-OBC processed data, the 
HTR3 and HTR5 modes, were determined using the known relation between Poisson 
level and observed raw count rate for these modes (Van der Klis 1989, 
Berger \&\ van der Klis 1994). In power spectra from OBC-processed data, 
the HER7 and HTR4 modes, the OBC dead time strongly 
affects the Poisson level (see e.g.\ Kuulkers et al.\ 1994a).
Moreover, the power spectrum of the Poisson noise in these power spectral 
data can no longer be considered constant at frequencies higher than 
$\sim$120\,Hz; it increases to higher frequencies (Tennant 1987). 
Most HER7 mode data had a Nyquist frequency of 128\,Hz or lower,
so this only had a significant effect in the HTR4 mode, which had a Nyquist 
frequency of 2048\,Hz. We used the Poisson level as a free parameter in the 
fits to the HER7 power spectral data. In the case of HTR4 we first binned the 
data to a time resolution of $\sim$4\,ms, and used the Poisson level as a 
free parameter in the power spectral fits. We refer to Appendix A for a more 
detailed discussion of dead time effects.

Whenever a power spectral component was not clearly present in the 
averaged power spectra, we determined a 1$\sigma$ upper limit to the 
rms amplitudes of this component by fixing its other parameters
(while keeping the parameters of the other observed components free).
The values of the fixed parameters were determined by evaluating
the observed behaviour (using only the HTR-mode data) of the corresponding 
components when they were clearly detected, and are discussed below.
\begin{description}
\item VLFN is present in all branches. Since there is no simple dependence of 
the VLFN power-law index as a function of the position in the Z 
(Section~4.2.2) we fixed it to its average observed value when 
determining an upper limit, i.e.\ 1.5.
\item LFN is observed to be only present in the HB and the upper part of the 
NB. Upper limits where therefore only determined for $S_{\rm Z}$ 
(see Section~4.2.1) up to $\sim$1.35. No clear dependence of the LFN 
power-law index and cut-off frequency on position in the Z 
is found (Section~4.2.3); the fixed values when upper limits were determined 
of these parameters were, therefore, $-$0.5 and 2\,Hz, respectively.
\item HFN is also present in the whole Z, but no clear dependence of the 
HFN cut-off frequency as a function of the position in the Z can be found 
(Section~4.2.4). In cases when an upper limit was determined the cut-off 
frequency was fixed at $\sim$77.
\item The HBO centroid frequency clearly depends on position in the HB 
(Section~4.2.5, see also Penninx et al.\ 1990). The power spectral fit 
results from HTR3-mode data on 1985 day 258/259 (Table 3a) indicate the 
following linear relation between the centroid frequency and 
$S_{\rm Z}$: $\nu_{\rm HBO}\,(Hz) = 24.0 + 12.5\,{\rm S}_{\rm Z}$. This 
relation was used when an upper limit on the HBO was determined;
for the HBO FWHM no clear dependence on the position in the Z is found 
(Section~4.2.5) and we then fixed it at 3.
\item A clear HBO harmonic was only detected on a few occasions 
(see Table 4, later). 
Whenever the fundamental HBO was clearly detected we determined an upper 
limit on its harmonic. In these cases the harmonic centroid frequency and 
harmonic FWHM were fixed to twice the value of the fundamental frequency and 
fundamental FWHM, respectively, whenever the source was in the HB. Recently, 
QPO between 50 and 62\,Hz were found in the upper and middle part of the NB 
of GX\,17+2 (Wijnands et al.\ 1996a). We regard this as the HBO harmonic 
(see Section~6.2.2). Between $S_{\rm Z}$ values of 1--1.75 we determined rms 
amplitude upper limits for such QPO using fixed values for the centroid 
frequency of 60\,Hz and FWHM of 5\,Hz. 
\item Upper limits to the NBO rms amplitude were determined in the entire NB. 
Average values of 3\,Hz and 7\,Hz were then used for the fixed centroid 
frequency and FWHM, respectively.
\item Upper limits to the FBO rms amplitude were determined in the entire FB. 
We then fixed its centroid frequency and FWHM to the observed value, 
i.e.\ 16\,Hz and 7\,Hz, respectively.
\end{description}

\section{Spectral and timing behaviour}

\subsection{Broad-band spectral analysis}

The CD of all HER observations after 1983 is shown in Fig.~1a.
In Figs.~1b and c we display the corresponding HIDs, 
with the 1986 data, which had pointing problems, left out.
The CD and HIDs of the 1983 day 215/216 data can be found in 
Figs.~2a--c. The data points in these figures
have been connected to show more clearly the presence of the FB.
It appears that the 1983 Z track is at higher soft and hard colours than that of
the other observations. The overall
intensities are comparable.
CDs of other sources show similar shifts between data for a certain
period in 1983 and later data. This was observed in
the Z sources GX\,5--1, Cyg\,X-2, GX\,349+2, and the 
atoll sources 4U\,1636--53 and Ser\,X-1 (see Kuulkers et al.\ 1996, and 
references therein). Since all sources show the same effect, the shift is probably 
instrumental.

All data on GX\,17+2 obtained after 1983 fall within a single Z track in the CD.
We can exclude shifts in the soft and hard colours at the level of $\sim$2\%\
(determined by examining the width of the branches).
We therefore conclude that the Z track of GX\,17+2 in the CD is stable to within 
$\sim$2 per cent and does not show evidence for secular motion in the CD for observations
separated by $\sim$1 day up to $\sim$1.5 years.

The HIDs of GX\,17+2 seem to be also quite stable. However, one data set 
(1985 day 259) is placed at $\sim$8 per cent higher intensities than the other 
data. Since instrumental changes between the 1984 day 249/250 and the 1985 
day 258/259 data can only be expected to contribute at most up to 3 per cent in 
intensity (see Kuulkers \&\ van der Klis 1996) we think this shift in 
intensity is probably intrinsic to the source. More observations are needed 
to verify if this effect is real and how it relates to other source properties.

As can be seen in Figs.~1a--c and 2a--c, GX\,17+2 was observed in all three 
spectral branches (see also Table~1). The HB is clearly not horizontal in 
the CD, but nearly vertical, as noted previously (e.g.\ Hasinger \&\ van 
der Klis 1989); in the other diagrams it has different orientations. Due to 
an unfortunate choice of HER7 energy boundaries when the 1986 observations 
were performed, the FB is positioned close to the NB (Fig.~4a). This also 
caused the FB to be oriented above the NB in Fig.~2b, while it falls along 
the NB in Fig.~2c. 

\subsection{Power spectral analysis}

\subsubsection{Introduction}

We studied the power spectral components as a function of position along the 
Z. We used a one-dimensional parameter, $S_{\rm Z}$ to measure curve length 
along the Z (Hertz et al.\ 1992, Dieters \&\ van der Klis 1996).
The scale in the CD was fixed by assigning $S_{\rm Z}$ values of 1 and 2 
for the HB/NB apex and the NB/FB apex, respectively, in the CD of Fig.~1a. 
The same scaling was also used for the 1983 day 215/216 observations.
We note that during the first part of the 1985 day 258/259 
observation (16:15--03:13 UT) we could not estimate $S_{\rm Z}$ from the CD, 
since data were only available in one broad low energy band. However, 
since the position in the HB is dependent on intensity
we could determine $S_{\rm Z}$ indirectly. For the second part of the 1985 day 
258/259 observation we had sufficient data to determine the dependence of 
intensity on $S_{\rm Z}$ in the relevant channel. We used this relation to 
determine $S_{\rm Z}$ values in the first part of the observations. 
Using the above defined scaling we find that 
$S_{\rm Z}$ ranges from $\sim$0 
in the upper part of the HB, to $\sim$4.5 in the upper part of the FB.
In Figs.~3a--c we show average power spectra from the different branches 
(HB, NB and FB). The various spectral components are indicated. 

In Tables~2, 3a and 3b we give the results of the power spectral fits as a 
function of $S_{\rm Z}$ for the 512\,s and 16\,s data segments, respectively. 
In Table~4 we give the results for the HBO harmonic (see Section 4.2.5).
Upper limits (1$\sigma$) have been determined on the rms amplitudes whenever a component
was not clearly present in the average power spectra. The range in S$_{\rm Z}$
where these upper limits where determined and the values to which the other parameters
of that particular component were fixed are described in Section~3.

In Figs.~4a--l most of the power spectral parameters have been plotted vs.\ 
$S_{\rm Z}$. In the next sections we discuss the results for each power 
spectral component. The 1985 day 258/259 HER7 and HTR4 data are not used in 
our discussion since they have different energy bands than the other data 
(see Table~1); only for the discussion of the HBO FWHM and frequency we 
include the HTR4 data, since these parameters are approximately independent 
of the energy (Penninx et al.\ 1990, Penninx et al.\ 1991, Lewin et al.\ 1992).

\subsubsection{Very-low-frequency noise (VLFN)}

The VLFN was detected in all spectral branches,
consistent with previous observations.
The VLFN rms and its power law index are plotted vs.\ $S_{\rm Z}$ in 
Figs.~4a and b, 
and Figs.~4c and d, for the 512\,s FFTs and 16\,s FFTs, respectively.
We did not include the 1986 day 093/094 data, which had
VLFN rms values between 5 and 6 per cent (Table 2), while in the other 
observations 
in the lower part of the NB the rms is around 1 per cent. We attribute this 
to the intensity variations arising from the changing collimator response on 
time scales longer than $\sim$100\,s, which causes leakage of power 
from lower frequencies into the passbands of the power density 
estimators with nominal frequencies in the VLFN frequency range (see 
Deeter 1984). The power law index during the 1986 093/094 observations is 
about 2, consistent with low frequency leakage.

The changing collimator response only affects frequencies below 0.01\,Hz. 
This is clear from Figs.~4c and d, where the 1986 day 093/094 VLFN components, 
here measured only down to 0.01\,Hz, fall along the other data points in both 
figures.

The VLFN rms is lowest ($\sim$0.5 per cent) in the upper part of the HB and the 
upper part of the NB. It shows a local maximum in the middle part of the HB 
of $\sim$1 per cent. From the upper part of the NB to the lower part of the 
NB the rms increases to $\sim$1.5 per cent. From the 16\,s FFTs we infer that 
the VLFN strength continues to rise from the lower part of the FB to the 
upper part of the FB, up to at least 3 per cent.

In the lower part of the NB the power law index
has values of $\sim$1.5. In the upper part of the FB
the power law index is about 2, as expected from leakage effects from slow
(source) intensity variations (see above). 

\subsubsection{Low-frequency noise (LFN)}

The LFN is peaked (see e.g.\ Fig.~3a) as is clear from the negative 
LFN power law index values between $-$1.5 and 0, with cut-off 
frequencies between 1 and 3\,Hz. In Figs.~4e, f and g we 
show the LFN rms, power law index and cut-off frequency, respectively, 
as a function of S$_{\rm Z}$. The rms of the LFN 
decreases from $\sim$3 per cent in the upper part of the HB to $\sim$1.5 per 
cent in the upper part of the NB (see also Tables~2 and 3a). 

The low rms and the large uncertainty in the LFN cut-off frequency in the 
upper part of the NB indicate that the LFN vanishes in this 
part of the NB. In the rest of the NB and in the FB no evidence for LFN was 
found.

\subsubsection{High-frequency noise (HFN)}

The HFN is also present in all branches, see Figs.~4h and i for the 512\,s 
FFTs, and Table 3a and 3b for the 16\,s FFTs. Like the VLFN, the HFN
is not correlated with any of the other power spectral components. The rms 
strength clearly decreases from the upper part of the HB to the middle part 
of the NB from $\sim$5.3 per cent to $\sim$2.5 per cent. From there onwards 
it remains between 2 and 3 per cent.  

The cut-off frequency is not well determined in many of the power spectra, 
mainly due to the relatively low Nyquist frequencies (64\,Hz) in the 1984 day 
249/250 and 250/251 data. Whenever it could be determined we found values 
between 10 and 50\,Hz.

\subsubsection{QPO: HBO, NBO and FBO}

HBO were only found in the 1985 day 258 data in the upper part of the HB 
(Tables~2, 3a and 3b, Fig.~3a). There are indications of a very weak HBO component 
in the expected frequency range in the 1984 day 249/250 data, but this component 
is not significant. The QPO frequency increases from $\sim$23 to $\sim$28\,Hz 
from the upper part of the HB to the middle part of the HB, while the FWHM and 
rms decrease from $\sim$7 to $\sim$2\,Hz and $\sim$2.5 per cent to $\sim$1 per cent, 
respectively. The HBO disappear below $\sim$1 per cent rms 
in the lower part of the HB.

A harmonic to the HBO peak with a frequency consistent with twice the 
HBO frequency is seen at the lowest $S_{\rm Z}$ values (see Table~4, Fig.~3a).
The harmonic has a FWHM between 10 and 20\,Hz and an rms of about 2.5 per cent.
The harmonic is about as strong as the HBO itself; the data do not allow
to distinguish between the harmonic peak being the same width, or 
twice as wide, as the main peak (see also Table~4).

NBO with frequencies between 5 and 7\,Hz were only detected in the middle and 
lower part of the NB (Fig.~3b), with an rms between 
1.5 and 2.5 per cent, and FWHM between 2 and 6\,Hz (Figs.~4j--l).

In the 1986 day 093/094 data we found evidence for FB QPO (FBO) in the lower 
part of the FB (Fig.~3c), with a centroid frequency of $\sim$16\,Hz, FWHM 
of $\sim$7\,Hz and rms of $\sim$2 per cent.

\section{Burst analysis}

\subsection{New bursts}

In Figs.~5a--e we show the light curves of all observations (except 
1986 day 093/094) at 5\,s time resolution. In Figs.~5c and d the smooth 
overall variations are interrupted by two bursts, which were already reported 
by Sztajno et al.\ (1986). In Figs.~5a--e we indicated which spectral
branch the source was in at each instant.

As noted in the previous section, the pointing during observation 1986 day 
093/094 was only stable to within 5\,arcmin. In Fig.~6 we 
display this data set (Fig.~6a, 5\,s time resolution) together with 
a soft-colour curve (Fig.~6b, 200\,s time resolution).
Although variations due to the varying collimator 
response are visible, one can also see that the variations in the FB are more 
irregular than in the NB.

Two sharp peaks seem to disturb the relatively smooth variations 
in the light curve of the 1986 day 093/094 data ($\sim$36\,000 and 
$\sim$104\,000\,s after the start of the observations). We investigated 
these peaks and found that they are X-ray bursts which had been previously 
overlooked. The housekeeping data ruled out a solar particle origin 
(see also e.g.\ Kuulkers et al.\ 1995).
In Figs.~7a and b we display light curves of these bursts at a higher time 
resolution (0.5\,s); plotted is the total HTR5 count rate,
corrected for dead time and background. The second burst straddled two 
{\it EXOSAT} observation segments and is therefore interrupted for $\sim$200\,s. 
The housekeeping monitoring during this time interval still provided 32\,s 
resolution qualified event rate data with which we overlayed the HTR5 light 
curve of this burst.

The bursts are not related to the changes in the pointing 
of the satellite. The events rise within 1\,s, which
is much faster than the pointing variations (typically $\sim$100\,s), and 
show clear hardness variations, which can not be caused by pointing 
variations. In Figs.~7c and d we display the 4.7--19.9\,keV/1.2--4.7\,keV 
hardness ratio curves of the two bursts. The bursts display 
spectral softening during their decay.

\subsection{Burst properties}

The first of the two new bursts started at 1986 day 093 14:04:57 (UT) and 
lasted for $\sim$100\,s, while the second started $\sim$19\,hr later at 
1986 day 094 9:05:40 (UT) and lasted $\sim$300\,s. No other bursts occurred 
in between these two during times that the OBC was operational.
We checked whether bursts occurred when the OBC was not operational using the 
housekeeping data, and found none. Since the collimator response changed on 
time scales of $\sim$100\,s during these bursts, it was not possible to 
accurately determine the standard X-ray burst parameters (see Lewin et al.\ 
1993) $\alpha$ (ratio of the average persistent flux to the time-averaged 
flux emitted in the bursts) and $\gamma$ (ratio of the mean persistent 
pre-burst flux and the net peak burst flux). However, it is clear that 
$\alpha$ was very large (larger than 1000) and $\gamma > 1$.
Spectral data (HER2) were only available during the second burst, but because 
the burst was interrupted and was observed with an uncertain, varying 
collimator response, we did not analyze these data.

To investigate in which branch the source was when the 1986 bursts and the 
bursts detected by Sztajno et al.\ (1986) occurred,
we indicated the position of the soft and hard color values in the CD
at the time the bursts occured with arrows in Fig.~1. We found 
that the small 1984 day 250 burst (I) occurred when GX\,17+2 was in the upper 
part of the NB, while the other bursts (1985 day 232: II; 1986 day 093: III; 
1986 day 094: IV) occurred in the middle part of the NB or near the NB/FB apex. 
Although none of the bursts occurred in the HB,
we note that GX\,17+2 is only occasionally found in the HB (see Section~6.1).
Moreover, using the total time spend in each branch as observed 
with EXOSAT, we found that the occurrences of the four EXOSAT burst 
are consistent with being independent of branch position.
We also found no clear relation between burst 
duration and position in the Z.

\subsection{Pulsations during the bursts?}

We searched for periodicities in the bursts. We followed the same procedure 
for searching pulse periods as Vaughan et al.\ (1994). We transformed 16\,s, 
256\,s and 64\,s data segments for the 1984, 1985 and 1986 bursts, 
respectively. We also averaged the power spectra of 64\,s data segments of 
the two 1986 bursts, and of these bursts together with the 1985 burst. 
We found no evidence for a periodic signal above the 99 per cent confidence 
detection level. The 99 per cent confidence upper limits to the amplitude of a 
sinusoidal modulation are given in Table~5. The best limits are 
$\sim$1 per cent for the 0--256\,Hz frequency range, and $\sim$1.5 per cent for the 
256--512\,Hz frequency range.

\section{Discussion}

\subsection{Two groups of Z sources}

GX\,17+2 and the other Z sources form a group with similar behaviour, 
i.e.\ three branches with correlated fast timing behaviour 
(Hasinger \&\ van der Klis 1989). However, it has been shown that GX\,17+2 
shares properties with Sco\,X-1 and GX\,349+2, which are distinct from 
GX\,5--1, Cyg\,X-2 and GX\,340+0 (Hasinger \&\ van der Klis 1989,
Penninx et al.\ 1991, Kuulkers et al.\ 1994a). 
The latter sources, for example, show long-term variations in the position 
of the Z pattern in the CD by typically $\sim$7 per cent 
(which are recurrent in position and shape in the case of Cyg\,X-2, see 
Kuulkers et al.\ 1996) and dipping behaviour in the
FB, while the former sources show a more stable Z pattern and flares in 
intensity in the FB. In this paper, we showed that in the CD the position of 
the Z track of GX\,17+2 is stable to within 2 per cent, which is in accordance with
this. 

It has been suggested that these differences are caused by a difference in 
orbital inclination angle (Kuulkers et al.\ 1996, and references 
therein). Optical observations of Cyg\,X-2 (Cowley, Crampton \&\ Hutchings 
1979) and Sco\,X-1 (Crampton et al.\ 1976) indicate orbital inclinations of 
65--75$^{\circ}$ and 
15--40$^{\circ}$, respectively. The fact that intensity decreases
in the FB of the Cyg-like sources is interpreted as obscuration of the 
central emission by the inner disk when it swells up as the 
luminosity approaches the Eddington limit.
Orbital inclination also plays a key role in suggested explanations for: (i) the 
secular variations in the shape and position in the CD and HID of the Z 
pattern of the Cyg-like sources (e.g.\ \mbox{Kuulkers} et al.\ 1994a, 1996), 
(ii) the shortness of the FB in GX\,5--1 (Kuulkers \&\ van der Klis 1995) 
and (iii) the peculiar QPO found in the FB of Cyg\,X-2 
(Kuulkers \&\ van der Klis 1995). 
For example, long term variations may be caused by changes in 
our view of the inner regions due to changes in the geometry
of structures near the orbital plane (such as accretion stream and outer disk,
see Kuulkers et al.\ 1994a, 1996, see also Wijnands, Kuulkers \&\ Smale 1996b). 

Several other properties (described below) in Z sources also appear to 
correlate with the division into two groups, but have not (yet) been 
interpreted in terms of orbital inclination differences. 

\begin{description}
\item Peaked LFN has been observed in GX\,17+2 and Sco\,X-1 (see e.g.\ 
Hasinger \&\ van der Klis 1989), while all three Cyg-like sources show 
non-peaked LFN. Atoll sources also show peaked and non-peaked noise 
(the so-called atoll source HFN). It has been suggested that atoll source 
HFN is similar to Z source LFN (van der Klis 1994). The different HFN 
shapes in atoll sources could perhaps be a manifestation of 
similar effects to those seen in Z source LFN.
\item A vertical (short) HB in the Sco-like sources has been seen,
versus a horizontal HB (with an upward bend at the lowest $\dot{M}$) in the 
Cyg-like sources.
\end{description}
The following property also appears different between the two groups, 
but seems hard to fit in with the orbital inclination framework:
GX\,17+2 and Sco\,X-1 are only occasionally found in the HB, 
and GX\,349+2 never, while GX\,5--1, Cyg\,X-2 and 
GX\,340+0 are frequently found in the HB, but only occasionally in the FB. 
Moreover, the Sco-like sources spend a much longer time scale in the FB
($>$12\,hr) than the Cyg-like sources ($<$2\,hr).
Since there is a good evidence that $\dot{M}$ increases from the NB to
the FB this implies that, on average, the Sco-like sources have a higher 
$\dot{M}$ than the Cyg-like sources. It is hard to understand how
this could have its origin in the inclination at which we view the sources,
unless viewing geometry affects the $\dot{M}$ level at which branch transitions 
occur in the CD and HIDs. Careful study of fast timing properties as a 
function of $S_{\rm Z}$ can in fact show whether this is the case, but the 
available data are not sufficient to make a strong case either way.

Recently, Psaltis et al.\ (1995) suggested the magnetic field strength of the 
neutron star to be the origin of the differences in the shape of the Z track 
in the Cyg-like and Sco-like sources (the magnetic field being lower in the 
latter sources). From their radiative transfer 
calculations they infer that subtle differences in the magnetic fields 
are able to explain the observed differences in the shapes of the 
Z pattern. It is not clear whether a difference in neutron star magnetic 
fields can explain the shifts in the Z track position, the apparent
difference in mean $\dot{M}$ between the two groups, or the 
differences in LFN shape, but it does seem likely that differences in HBO 
properties between the two groups can be explained by it.

\subsection{Timing behaviour}

\subsubsection{Very-low-frequency noise}

The VLFN was detected in all branches. 
Near the HB/NB vertex it was weakest ($\sim$0.5 per cent rms), while in the 
FB it was strongest (rms of up to at least $\sim$3 per cent rms).
We found power law indices between $\sim$0.8 and $\sim$2.

The VLFN of Z sources has been investigated in detail in several other studies 
as a function of $S_{\rm Z}$ (\mbox{GX\,5--1}: Kuulkers et al.\ 1994a, 
{\it EXOSAT} argon; Sco\,X-1: Dieters \&\ van der Klis 1996, {\it EXOSAT} 
xenon; Hertz et al.\ 1992, {\it Ginga}; GX\,340+0: Kuulkers \&\ van der Klis 
1996) or as a function of rank number (GX\,5--1: Lewin et al.\ 1992, 
{\it Ginga}; Cyg\,X-2: Hasinger et al.\ 1990, Wijnands et al.\ 1996c, 
{\it Ginga}; Hasinger 1993 [private communication], {\it Ginga} and 
{\it EXOSAT} argon; GX\,340+0: Van Paradijs et al.\ 1988a, \mbox{{\it EXOSAT}} 
argon; Penninx et al.\ 1991, {\it Ginga}; GX\,17+2: Penninx et al.\ 1990, 
{\it Ginga}). Hasinger \&\ van der Klis 
(1989) investigated {\it EXOSAT} data of the six Z sources as a function of 
HB, NB and FB. Comparing the {\it EXOSAT} argon and the {\it Ginga} results 
(both instruments provided roughly the same energy range), we find that the 
VLFN is lowest in the upper part of the NB, near the
HB/NB vertex. The minimum strength is similar for all the 
sources, i.e.\ near 0.5 per cent rms. The increase in VLFN from the upper 
part of the NB to the lower part of the FB is strongest in GX\,5--1
(Kuulkers et al.\ 1994a) and Cyg\,X-2 (Hasinger 1993, private communication).
The VLFN rms near the NB/FB vertex is $\sim$4 per cent in GX\,5--1 
(Kuulkers et al.\ 1994a), while it is about 2 per cent for 
GX\,340+0 (Van Paradijs et al.\ 1988a, Penninx et al.\ 1991), GX\,17+2 
(this paper) and Sco\,X-1 (Hertz et al.\ 1992). 
The sources GX\,5--1 and GX\,17+2 show evidence for a maximum of $\sim$1 per 
cent in the middle part of the HB  (Kuulkers et al.\ 1994a and this paper, 
respectively). As shown by Dieters \&\ van der Klis (1996), the VLFN is 
stronger in the 5--35\,keV xenon data (see also Stella et al.\ 1987), 
i.e.\ at higher energies.

The power law indices for GX\,340+0 (Penninx et al.\ 1991), GX\,17+2 
(Penninx et al.\ 1990, this paper) and Sco\,X-1 (Hertz et al.\ 1992, 
Dieters \&\ van der Klis 1996) show more or less the same behaviour as a 
function of $S_{\rm Z}$. The index is low ($\la$1) in the HB and 
upper part of the NB, while it increases to $\sim$2 near the NB/FB vertex. 
In the FB VLFN power law indices around 2 have been found (note that the true 
values could be even higher, due to the leakage of power in the VLFN 
frequency range). GX\,349+2 (Hasinger \&\ van der Klis 1989) shows behaviour 
consistent with this. GX\,5--1 (Kuulkers et al.\ 1994a) shows a clearly 
opposite behaviour. The power law index remains at about 2 along the HB to 
the middle part of the NB, and then drops to $\sim$1.4 in the lower part of 
the FB. 

In the model of Lamb (1989) the increase of VLFN from the lower part of the 
NB into the FB is explained by instabilities in the accretion flow. When 
$\dot{M}$ reaches the Eddington limit radiation pressure becomes important. 
Accretion onto the neutron star becomes highly unstable at this stage, and 
therefore increases the VLFN strength.

\subsubsection{Horizontal branch QPO}

In the upper/middle part of the HB we found HBO with frequencies between 24 
and 28\,Hz, FWHM between 1 and 8\,Hz, and fractional rms amplitudes between 
1.5 and 2.5 per cent. The HBO decreased rapidly in strength from the upper 
part of the HB to the middle part of the NB. We found evidence for a second 
harmonic, with a frequency consistent with twice the HBO frequency,
and with similar strength as the first harmonic.
These results are similar to those reported by Stella et al.\ (1987) and 
Langmeier et al.\ (1990) for the {\it EXOSAT} data. 
The range of the HBO frequencies falls within the range (18--30\,Hz from 
upper to lower part of the HB) found by Penninx et al.\ (1990) in {\it Ginga} 
data. 

More recently, variable-frequency QPO have been found on the NB of GX\,17+2 
(Wijnands et al.\ 1996a). These authors interpreted these QPO as being 
(fundamental) HBO which persists on the NB. However, if we extrapolate the 
(linear) relation 
between HBO fundamental centroid frequency and $S_{\rm Z}$, using our 
HTR-mode EXOSAT data, we find a frequency of $\sim$36.5\,Hz at the HB/NB 
vertex ($S_{\rm Z}$=1), which is near the maximum value found in the 
lower part of the HB in the {\it Ginga} data (see above). A similar conclusion 
can be derived from the {\it Ginga} power spectral parameters for the
HBO (Penninx et al.\ 1990). Extrapolating the (linear) relation between 
HBO fundamental centroid frequency and either {\it Ginga} MPC3 colour or 
{\it Ginga} PC colour 
to $\sim$60\,Hz give HB/NB vertex colour values which are way off from 
the expected place of the vertex in the {\it Ginga} CD and HID.
We therefore interpret the variable-frequency QPO as being a harmonic of 
the HBO, unless the HBO fundamental centroid frequency changes abruptly
(by 15--20\,Hz) near the HB/NB vertex. The linear change of the HBO harmonic 
centroid frequency (see Penninx et al.\ 1990) as a function of either 
{\it Ginga} MPC3 colour or {\it Ginga} PC colour, extrapolated to the expected 
HB/NB vertex in the {\it Ginga} CD and HID are consistent with this 
interpretation.

The other Z sources in which HBO have been found, GX\,5--1 (Lewin et al.\ 
1992, Kuulkers et al.\ 1994a), \mbox{Cyg\,X-2} (Hasinger 1987a, 
Hasinger et al.\ 1990, Wijnands et al.\ 1997) and GX\,340+0 (Penninx et al.\ 1991,
Kuulkers \&\ van der Klis 1996) show 
HBO properties that are somewhat different from those of GX\,17+2.
The maximum HBO frequencies are much higher in the former sources 
(up to 55\,Hz, with rms amplitudes between 2 and 6 per cent). 
In GX\,5--1 a second harmonic with 
an rms amplitude roughly half that of the first harmonic was found (Lewin 
et al.\ 1992, Kuulkers et al.\ 1994a); similar harmonics are reported 
in the HBO of Cyg\,X-2 (Hasinger 1987a, Focke 1996, Wijnands et al.\ 1997) and
GX\,340+0 (Kuulkers \&\ van der Klis 1996).

The HBO are thought to be caused by magnetospheric gating (Alpar \&\ Shaham 
1985, Lamb et al.\ 1985, Ghosh \&\ Lamb 1992), at
the beat-frequency between the neutron star spin frequency ($\nu_{\rm NS}$) 
and the Kepler frequency at the inner disk edge ($\nu_{\rm K0}$). 
This beat-frequency model requires inhomogeneities in the accretion flow, 
whose presence has as a natural consequence the production of LFN.
The radius of the inner edge of the disk depends on $\dot{M}$, the magnetic 
moment $\mu$ and the mass $M$ of the compact object
(Ghosh \&\ Lamb 1992). For n-fold symmetry in the magnetic field pattern, 
this leads to (see Lamb et al.\ 1985, Ghosh \&\ Lamb 1992):
\begin{equation}
\nu_{\rm HBO} = n (C \dot{M}^{\alpha}\mu^{\beta}M^{\kappa} -\nu_{\rm NS}),
\end{equation}
where $n$ is the harmonic number, $C$ is a constant, and $\alpha$, $\beta$ 
and $\kappa$ are exponents which depend on the particular inner disk model 
considered; $\alpha>0$, $\beta<0$, whereas $\kappa$ can be either positive or 
negative. There are various ways to produce the $\sim$factor 2 lower HBO 
frequency range in GX\,17+2 as compared to the other Z sources: lower the 
overall mass-accretion rate, increase the magnetic field strength or the mass 
of the neutron star, either increase or decrease
the spin period of the neutron star depending on the value of $\kappa$, or 
lower the harmonic number $n$ (if $n$$>$1). Of these possibilities, differences 
in $\dot{M}$ and $M$ are unlikely. $\dot{M}$  is probably the same at the 
apexes for all Z sources (e.g.\ Lamb 1989, Hasinger et al.\ 1990).
(But perhaps the inclination influences the position of the apexes in an 
unknown way, see also Section 5.2.) Unless $\kappa$ is large, neutron star 
masses are not likely to be sufficiently different to explain 
a factor of 2 in the maximum HBO frequency.
A closer match between $\nu_{\rm NS}$ and $\nu_{K0}$ in GX\,17+2 
than in the other Z sources would of course also lead to a lower
HBO frequency. However, in that situation the {\it range} in HBO frequency, 
which is determined by the range in
$\nu_{\rm K0}$, would still be expected to be similar to 
that in the other sources, whereas it is lower:
$\sim$12\,Hz in GX\,17+2 versus $\sim$35\,Hz in GX\,5--1 and Cyg\,X-2.

Since the maximum HBO frequency in GX\,17+2 is about a factor of two lower 
than the maximum HBO frequency in Cyg\,X-2, GX\,5--1 and GX\,340+0, we now 
explore the possibility that the HBO frequency seen in GX\,17+2 is 
dominated by the fundamental (first harmonic) of the field symmetry ($n$=1) and 
the HBO frequency seen in the latter sources by the second harmonic ($n$=2). 
This could be true if in GX\,17+2 the radiation from one magnetic pole at the 
neutron star surface dominates that from the other magnetic pole,
whereas in the other sources both poles contribute approximately the same
amount of radiation. The less luminous magnetic pole in GX\,17+2 is expected, 
if its luminosity is not negligible, to give rise to a second harmonic with 
centroid frequencies in a similar range as the HBO in the other sources, as 
actually observed in GX\,17+2. The second harmonic seen in GX\,5--1 and 
GX\,340+0 is weak compared to that in GX\,17+2 and in our interpretation due 
to deviations of the HBO signal from being sinusoidal. If the above 
described model is right, we predict the presence of a subharmonic
(due to the difference between the emission from the two poles) in the 
power spectra of GX\,5--1, Cyg\,X-2 and GX\,340+0, while
it should be absent in the power spectra of GX\,17+2.

Differences in the radiation from the two magnetic poles could be caused by 
asymmetries in the field geometry, which lead to different rates of accretion 
onto, or different emission characteristics from, the two
poles. These asymmetries would also lead to a difference in the
surface area of the two polar caps. Different polar cap areas have also been 
deduced from pulse analyses in several pulsars (e.g.\ Leahy 1991, 
Bulik et al.\ 1992). 

We note that in a situation with unequal polar cap areas type~I bursts might 
have a higher chance to occur, as on the larger of the two polar caps 
accretion per unit area would be less, and therefore on that pole
the conditions for a type\,I burst would be more favourable 
than for two equal-area polar caps. The presence of bursts and the lower 
maximum HBO frequency might therefore be explained together in this model.

We shall for illustrative purposes consider one simple asymmetry in field 
geometry, namely that which arises when the (dipole) magnetic field center 
does not coincide with the neutron star center, but is shifted.
In Figs.~8a and b we plotted a magnetic dipole field in which the neutron 
star center coincides with the magnetic field origin (Fig.~8a) and one in 
which the neutron star origin is somewhat shifted (Fig.~8b). In these figures 
we have indicated the polar caps as expected from the last closed field line 
in a simple dipole plus spherical magnetosphere geometry.
It is known that deviations from this geometry will
occur, especially near the inner disk edge, 
because of the magnetic field of the accretion disk (e.g.\ Ghosh \&\ Lamb 1979, 
Spruit \&\ Taam 1990), but we ignore this here. 
We computed the areas of the two polar caps by assuming magnetospheric radii 
between 15 and 100\,km (which is roughly the range expected in Z sources, see 
Ghosh \&\ Lamb [1992]), and offsets of the magnetosphere origin with respect 
to the neutron star origin between 0 and 5\,km.
The neutron star was assumed to have a radius of 10\,km.
In Table~6 we give the results.
As can be seen in this table, a slight offset of the magnetic dipole 
has a large effect on the polar cap areas. 
The ratio of the two polar cap areas already exceeds 2 for offsets of only 
$\sim$1.5\,km. For small magnetospheric radii, a small offset will cause
the larger of the two polar caps to cover a considerable fraction of its 
hemisphere.

When $\dot{M}$ increases along the Z, the magnetospheric radius becomes 
smaller. This leads to a rapid increase in the polar cap areas, 
which might explain the fact that
the four bursts observed with {\it EXOSAT} occurred in the NB and FB, and not in the 
(lower $\dot{M}$) HB. However, somewhat dependent on the inner disk model
(see Ghosh \&\ Lamb 1992), accretion per unit area is usually still expected to 
increase when total $\dot{M}$ increases in our simple model, even on the 
larger of the two polar caps.

The presence of multipole components in the magnetic field is another
possibility to cause the required asymmetries between the accretion regions.
As multipole effects increase very strongly with decreasing radius, these
effects would be most prominent for small magnetospheric radius, i.e.,
at higher $\dot{M}$ or for lower overall magnetic field strengths. An overall
lower magnetic field in GX\,17+2 (as suggested by Psaltis et al.\ 1995, see
the discussion in Section~6.1) could also be the explanation of the
occurrence of bursts in this source, and the increase of $\dot{M}$ 
from HB towards NB might explain why no bursts have been found in the HB.

\subsubsection{Flaring branch QPO}

On one occasion we found evidence for FBO with a frequency, FWHM and rms 
amplitude of 16\,Hz, 7\,Hz and 2 per cent, respectively. Similar 
FBO in GX\,17+2 were reported earlier in {\it Ginga} data by Penninx et al.\ (1990) 
with frequencies up to 20\,Hz and rms up to 3 per cent, but not from {\it EXOSAT} data. 
Sco\,X-1 is the only other Z source for which similar FBO have been reported 
(see Dieters \&\ van der Klis 1996, and references therein). This kind of 
FBO merges smoothly with the NBO, and therefore probably has a common origin.
FBO, and the NBO, are believed to be due to oscillations in the optical depth 
of the radial flow in the inner disk region (Lamb 1989, 
Fortner, Lamb \&\ Miller 1989) 
or due to sound waves in the thick accretion disk (Hasinger 1987b, 
Alpar et al.\ 1992).

It was recently suggested that the strength of the NBO oscillations depends 
on the inclination at which we view the Z sources (Kuulkers \&\ van der Klis 
1995), stronger NBO occurring in the lower-inclination sources. The strength 
of the NBO in the lower NB and the fact that we observe FBO in GX\,17+2 are 
in accordance with this. It is not clear how this difference could be
explained in the Psaltis et al.\ (1995) model for the two groups of
Z sources.

\subsubsection{High-frequency noise}

We found that in GX\,17+2 the strength of the 
HFN decreases as a function of $S_{\rm Z}$ from the HB to the 
middle part of the NB from $\sim$5.3 per cent to $\sim$2.5 per cent. 
From the middle part of the NB to the FB the HFN is more or less constant at 
a level of $\sim$2.5 per cent. This is different from what is seen in Sco\,X-1
(Hertz et al.\ 1992, Dieters \&\ van der Klis 1996), and probably Cyg\,X-2
(Hasinger et al.\ 1990, Hasinger 1993, private communication). There the HFN 
is reported to decrease from the HB, to the NB, {\it into} the FB. 
None of these observations were obtained with the {\it EXOSAT} ME argon 
detectors, as our observations. The results for Sco\,X-1 and Cyg\,X-2 from {\it Ginga} 
(Hertz et al.\ 1992, Hasinger et al.\ 1990, Hasinger 1993, private 
communication) indicated values which were typically lower than $\sim$1.5 
per cent in the FB. It has recently been suggested that some of the HFN 
seen with the {\it EXOSAT} ME is instrumental, up to an rms strength of $\sim$2.3 
per cent (0.01--100\,Hz) in the argon data (Berger \&\ van der Klis 1994, 1997). 
This instrumental effect has an amplitude similar to the values found in the 
middle part of the NB and FB of GX\,17+2. We conclude that our data are 
therefore consistent with a decrease in HFN strength from the HB to the FB.

\subsection{Bursts in GX\,17+2}

We have detected two bursts in GX\,17+2
in the 1986 data. One burst was about 100\,s long, the 
other (which occurred $\sim$19\,hr later) lasted $\sim$300\,s.
Previously, Sztajno et al.\ (1986) had found two bursts in the 1984 and 1985
data of {\it EXOSAT}, which lasted $\sim$10\, and $\sim$300\,s, respectively. 
Bursts in this source have also been observed with other X-ray instruments: 
one with {\it Einstein} ($\sim$10\,s, Kahn \&\ Grindlay 1984) and four with 
{\it Hakucho} (two which lasted for $\sim$100\,s and two which lasted for 
$\sim$300\,s, Tawara et al.\ 1984). All bursts have large $\alpha$ values 
($\alpha$$>$1000) and 
$\gamma$ larger than 1 (i.e., the net peak burst flux is smaller than the 
average persistent flux). 

We found that the {\it EXOSAT} bursts
occurred in the NB and lower part of the FB, i.e.\ at near-Eddington mass 
accretion rates. None were found in the HB, i.e.\ when $\dot{M}$ is lowest,
although we can not rule out the possibility that this is a coincidence. 
The only other Z source that has shown X-ray bursts is Cyg\,X-2
(Kahn \&\ Grindlay 1984, Kuulkers et al.\ 1995, Wijnands et al.\ 1997,
Smale et al.\ 1996). The occurrence of these bursts was found to be independent
of branch position (Kuulkers et al.\ 1995, Wijnands et al.\ 1997).

Bildsten (1993, 1995) suggested that rotating patches of nuclear-burning 
material on the neutron star surface during X-ray bursts may produce regular 
pulsations (see also Schoelkopf \&\ Kelley 1991). We searched for such 
pulsations in all four {\it EXOSAT} bursts, but found none, with
upper limits to the modulation strength of $\sim$1 per cent (if the signal is 
sinusoidal). A similar lack of pulsations was found for the persistent 
emission (Vaughan et al.\ 1994b) and in a large sample of X-ray bursts from other 
sources (Jongert \&\ van der Klis 1996). 

From the fact that the two new EXOSAT bursts show spectral
softening during their evolution, we conclude that, like the bursts reported 
by Sztajno et al.\ (1986), they are type I X-ray bursts caused by 
thermonuclear flashes on the neutron star surface (Hoffman, Marshall \&\
Lewin 1978, see also Lewin et al.\ 1993 and references therein).
The bursts in Cyg\,X-2, although much weaker than those in GX\,17+2, 
are also thought to be due to thermonuclear flashes, since the burst 
properties resemble those of type~I bursts (see Kuulkers et al.\ 1995, 
Wijnands et al.\ 1997). However, no evidence for spectral softening was
found. A burst recently detected with RXTE in Cyg\,X-2 (Smale et al.\ 1996) 
is the first case of a bona fide type~I burst in this source.
The fact that two Z sources show type~I bursts indicates that the compact 
star in systems of this type is a neutron star.

High luminosity burst sources tend to show irregular bursting behaviour 
(Van Paradijs et al.\ 1988b). They show bursts after long and variable waiting 
times. This is most probably due to steady nuclear burning, which occurs at 
high mass accretion rate $\dot{M}$. It is unlikely that $\dot{M}$ is lower in 
GX\,17+2 and Cyg\,X-2 as compared with the other Z sources, as in the NB 
$\dot{M}$ 
is believed to be approximately Eddington in all Z sources (Lamb 1989).
If bursts do apparently occur at these high levels of $\dot{M}$
(see e.g.\ Taam, Woosley \&\ Lamb 1996), then it is 
not clear why bursts are not seen in the other Z sources as well. 
From the presence of HBO in Z sources one derives that they contain a neutron 
star with a non-negligible magnetic field. In Section~6.2.2 we have
discussed the possibility that the magnetic field strength might affect the 
probability to produce bursts in a Z source.

\section{Summary}

We found that the position of the Z pattern in the CD is not moving by more 
than a few percent between observations separated by from 1 day up to
$\sim$1.5 years. This is in accordance with the division of the Z sources 
in two groups, in which the Cyg-like sources (GX\,5--1, Cyg\,X-2 and 
GX\,340+0) show secular variations in their Z pattern in the CD, while the 
Sco-like sources (GX\,17+2, Sco\,X-1 and GX\,349+2) do not.

We have investigated the various power spectral components as a function of 
position in the Z, $S_{\rm Z}$, and placed their properties in the framework 
of the division of Z sources into two groups and discussed the hypothesis 
that this division is due to differences in viewing geometry. Many, but not 
all, source properties fit in this picture.

Two new X-ray bursts were found in GX\,17+2 lasting $\sim$100 and 
$\sim$300\,s, which came at an interval of $\sim$19\,hr.
These bursts, and also those seen by Sztajno et al.\ (1986), occurred in the NB
or in the lower part of the FB, i.e.\ presumably near Eddington accretion 
rates. No evidence for regular pulsations was found in the bursts.

We suggest that an asymmetric magnetic field, 
giving rise to different surface areas and emission characteristics of 
the magnetic pole areas at the neutron star
surface may simultaneously explain the occurrence of bursts in GX\,17+2, 
the fact that they were not seen at the lowest
accretion rates, the fact that the maximum HBO frequency in GX\,17+2 is lower 
by about a factor two than the maximum HBO frequency in Cyg\,X-2, GX\,5--1 
and GX\,340+0 near the HB/NB apex, and the fact that the observed range in 
HBO frequency in GX\,17+2 is about a factor of two smaller than that of 
Cyg\,X-2 and GX\,5--1. The asymmetry could for example be due to an
overall lower field strength, allowing the multipole component of the 
magnetic field to play a larger role due to the smaller radius of the 
atmosphere.

\section*{acknowledgements}

Stefan Dieters is gratefully acknowledged for his 
discussions. We also thank him and Frank Verbunt for comments on earlier
versions of this paper. This work was supported in part by the Netherlands 
Organization for Scientific Research (NWO) under grant PGS 78-277.
WHGL acknowledges support from NASA.

\appendix
\setcounter{equation}{0}

\section{Dead time and Poisson noise}

Dead time is a severe problem in X-ray astronomy (e.g.\ Lewin, Van Paradijs \&\ 
van der Klis 1988, 
van der Klis 1989). Once an X-ray photon
is detected, the detector remains ``dead'' for a certain or variable time.
During this ``dead time'' the detector is unable to register new incoming 
X-ray photons. The various dead time processes onboard {\it EXOSAT} have been 
described by Andrews (1984), Andrews \&\ Stella (1985) and Tennant (1987). 
When the detector is dead for a fixed amount of time after 
a photon has been detected, one speaks of a
fixed dead time. In the case the detector can only register one 
photon within a given time (or ``sample cycle''), 
the dead time is variable (i.e.\ depending on where the photon is detected 
within the sample cycle), and one speaks in terms of variable dead time.
Within the ME instrument (i.e.\ the detector, electronics and the OBC) 
both constant dead time (in non-OBC processed data, such as 
HTR3 and HTR5 observation modes) and variable dead time 
(in HTR4 and all HER observation OBC modes) processes occur
(Andrews 1984, Andrews \&\ Stella 1985, Tennant 1987, Berger \&\ van der 
Klis 1994).

Dead time processes have recognizable effects on power spectra 
(see Lewin et al.\ 1988, van der Klis 1989). 
Pure counting noise, in the absence
of dead time, has a white power spectrum with, in 
the case of the Leahy et al.\ (1983) normalization,
an average power of two. This white noise component is 
called Poisson noise and its level the Poisson level. The Poisson level is
modified by the dead time process. The larger the dead time, the lower the 
Poisson level. Also the dead-time modified Poisson noise is not 
necessarily white any more. 
The effect of the fixed dead time of HTR3 and HTR5 on the power spectrum has 
been studied by Berger \&\ van der Klis (1994, 1997).

In the case of variable dead time, the Poisson level is approximated as 
follows (e.g.\ van der Klis 1989):
\begin{equation}
P_{\rm Poiss} = 2(1-\mu\tau_{\rm sample}),
\end{equation}
where $\tau_{\rm sample}$ is the duration of the sample cycle, and
$\mu$ the observed count rate.
For the {\it EXOSAT} ME value of $\tau_{\rm sample}$, 1/4096\,s, this 
expression gives wrong results. For a so-called effective 
sampling time of 1/3569\,s (Andrews \&\ Stella 1985) an approximate match
with the data is found.

Recently we found (Kuulkers et al.\ 1994a) that for high observed count rates
(1500--2300\,cts\,s$^{-1}$) in the OBC processed HER7 data 
the fitted Poisson levels deviated from that expected from Eq.~A1.
Since the HER7 data of GX\,17+2 provided lower count rates 
(0--1500\,cts\,s$^{-1}$), we decided to analyse all four-channel HER7 data 
for the brightest LMXBs, i.e.\ Sco\,X-1, GX\,5--1, Cyg\,X-2, GX\,17+2 and 
GX\,3+1. The former four sources are all Z sources, while the latter
is an atoll source (see e.g.\ Hasinger \&\ van der Klis 1989). In Table~A1, 
we give an observation log of the four channel HER7 argon observations of 
these sources. We selected the power spectra according to observed intensity 
and determined the Poisson level by fitting the power spectra with the 
various source components as described in Section~3, with the Poisson level 
as a free parameter. This was done for each source individually. 
The results are given in Fig.~A1. In this figure we also plotted the Poisson 
levels as predicted from the linear relation Eq.~A1. It is clear that the
fitted Poisson levels deviate from those expected (especially at high count 
rates), as was already found by Kuulkers et al.\ (1994a).
It can be seen that the Poisson level deviations from the predicted linear 
relation in the various sources where they overlap
are the same. The deviation is therefore not an effect related to variability 
in the sources themselves.

As mentioned in Section 3, we rebinned the OBC processed HTR4 data to a time 
resolution of $\sim$4\,ms, i.e.\ similar to most of the HER7 data. In the 
power spectral fits of the HTR4 data the Poisson level was also taken to be a 
free parameter. The results of these Poisson level fits are also 
plotted in Fig.~A1. As can be seen in this figure, the HTR4 data fall along 
the HER7 data points, as expected from the fact that these data are processed 
in a similar way, by the OBC, to the HER7 data.

Although a linear deviation was sufficient to describe the GX\,5--1 data of 
Kuulkers et al.\ (1994a), they noted that the Poisson level could be better 
described as a quadratic function of observed intensity.
Here we find that a linear fit is unacceptable, and that 
our data are described by a quadratic function.
When fitting the data to such a function, the Poisson level as a function of
observed intensity, $\mu$, is as follows: 
\begin{equation}
P_{\rm Poiss~HER7} = 1.9959(19) - 5.89(4)\cdot 10^{-4}\mu + 3.04(13)\cdot 10^{-8}\mu^2,
\end{equation}
where we give the uncertainty in the last digit(s) in brackets. The fitted 
Poisson levels in the observed count rate range 100--900\,cts\,s$^{-1}$ are 
below the expected Poisson levels. This is consistent with a previous study 
by Rutledge (1993 private communication) from HER7 data of the Rapid Burster.

We note that it was possible to fit the following function to the data:
\begin{equation}
P_{\rm Poiss~HER7} = 2\left( 1+ \sum^n_{i=0} (-1)^i\frac{1}{i^2}(\mu\tau_{\rm sample})^i \right).
\end{equation}
We found that $\chi^2_{\rm red}$ was $\stackrel{<}{\sim}$1.0 for $n$=4.
In that case $\tau_{\rm sample}$ was found to be 1/3304\,s.

\vspace{2cm}

\noindent
\begin{center}
{\bf Figure Captions}
\end{center}
{\bf Figure 1: }\\
({\bf a}) {\it EXOSAT} X-ray colour-colour diagram and ({\bf b}, {\bf c})
hardness-intensity diagrams
of GX\,17+2. Soft colour is defined as the ratio of
the count rates in the 4.7--6.6\,keV and the 1.2--4.7\,keV  bands,
hard colour as the ratio of the counts in the 6.6--19.9\,keV and the
4.7--6.6\,keV bands.
The intensity (corrected for background, dead-time and collimator
response) is defined as the count rate in the 1.2--19.9\,keV band.
Different symbols refer to different observation periods, which are
given in the lower right part of ({\bf a}). Each point represents a 200\,s
average. Typical error bars are given in each frame.
HB (horizontal branch), NB (normal branch) and FB (flaring branch) are
indicated. In {\bf a} we give the position in the CD at the time
the different bursts (1984 day 250: I, 1985 day 232: II, 1986 day 093:
III, 1986 day 094: IV) occurred (see Section 5.2).\\
~\\
{\bf Figure 2: }\\
({\bf a}) Colour-colour diagram and ({\bf b}, {\bf c}) hardness-intensity
diagrams of GX\,17+2 of the 1983 data. The colours and intensity are as
defined in Fig.~1. The points are connected to show
more clearly the different branches, which are indicated. Each point 
represents a 200\,s average.\\
~\\
{\bf Figure 3: }\\
Three examples of average power spectra corresponding to different segments
in $S_{\rm Z}$. The corresponding fits are shown for clarity, and the power spectral
components are indicated.
({\bf a}) Power spectra for data of 1985 day 258/259 when GX\,17+2 was in the
upper part of the HB near $S_{\rm Z}$$\sim$0.1. Both the HBO and its harmonic 
can be seen, as well as the peaked noise component LFN.
({\bf b}) Power spectra for data of 1985 day 232/233 during the NB near
$S_{\rm Z}$$\sim$1.8. The NBO can be seen near 7.6\,Hz.
({\bf c}) Power spectra for data in the lower part of the FB during 1986 day 
093/094 near $S_{\rm Z}$$\sim$2.4. The FBO can be seen near 16\,Hz.\\
~\\
{\bf Figure 4: }\\
({\bf a})--({\bf l}) Properties of the various power spectral components 
as a function of position on the ``Z'', $S_{\rm Z}$. The different
symbols refer to different observation periods as indicated in Fig.~1a.
The 1983 day 215 data are indicated with {\small $\bigcirc$}.
The boundaries between the spectral branches are indicated.
({\bf a}) VLFN rms 512\,s FFTs, ({\bf b}) VLFN power law index 512\,s FFTs,
({\bf c}) VLFN rms 16\,s FFTs, ({\bf d}) VLFN power law index 16\,s FFTs,
({\bf e}) LFN rms 16\,s FFTs, ({\bf f}) LFN power law index 16\,s FFTs,
({\bf g}) LFN cut-off frequency 16\,s FFTs, ({\bf h})
HFN rms 512\,s FFTs, ({\bf i}) HFN cut-off frequency 512\,s FFTs,
({\bf j}) NBO frequency, ({\bf k}) NBO rms ({\bf l}) NBO FWHM.\\
~\\
{\bf Figure 5: }\\
{\it EXOSAT} ME light curves of 4 observation periods of GX\,17+2 at a time
resolution of 5\,s. Intensity is the dead-time-, background- and 
collimator-response-corrected count rate in the 1.2--19.9\,keV range. 
Indicated are the 
branches in which the source was during the observations. During two 
observation periods (1984 day 251/250, [{\bf c}], and 1985 day 232/233, 
[{\bf d}]) bursts can be seen which were already reported by
Sztajno et al.\ (1986). Start times for the different
observation periods can be found in Table 1.\\
~\\
{\bf Figure 6: }\\
{\it EXOSAT} ME light curve ({\bf a}) and soft colour curve ({\bf b}) of the
observation period 1986 day 093/094. Intensity is the dead-time and 
background corrected count rate in the 1.2--19.9\,keV range, while the soft 
colour is the ratio of the count rate in the 4.7--6.6\,keV and the count 
rate in the 1.2--4.7\,keV band. The intensities are not corrected for 
collimator response (see text). Two new bursts can be seen at 
($\sim$36\,000 and $\sim104\,000$\,s after the start of the observation).
The start time of the observation period can be found in Table~1.\\
~\\
{\bf Figure 7: }\\
High time resolution (0.5\,s) argon light curves of the
bursts which occurred during 1986 day 093 ({\bf a}) and 1986 day 094
({\bf b}), together with their colour (ratio of the count rates in the
4.7--19.9\,keV and 1.2--4.7\,keV bands) curves, [{\bf c}] and [{\bf d}],
respectively. During the 1986 day 094 burst the observations were interrupted 
for mode changes. The qualified event rate from the housekeeping data is 
therefore also plotted in {\bf b} (with symbol {\small $\bigcirc$}) to show the 
overall burst profile. \\
~\\
{\bf Figure 8: }\\
({\bf a}) A neutron star (radius 10\,km) with a symmetric dipole field. The 
spherical magnetosphere (indicated by a dotted line)
in this plot is taken to be at a radius of 30\,km. The last closed field 
line is indicated with a dashed line. The corresponding accretion polar cap 
on the surface of the neutron star is indicated with a fat line. ({\bf b}) 
Same neutron star with a shifted (by 5\,km) magnetic dipole field.
The upper pole now has a much larger area than the lower pole.\\
~\\
{\bf Figure A1: }\\
Observed Poisson levels for the power spectra of HER7 data
for several LMXBs as a function of the raw observed count rate.
The HTR4 data of GX\,17+2 have also been plotted.
The observed Poisson level was determined by adding it as an extra
component in the power spectral fits. The different
symbols refer to the different sources as indicated.
Error bars are on the order of the symbol sizes.
The predicted Poisson levels (straight line) and the quadratic fit
(dashed line) are included in the plot.

\bsp 

\label{lastpage}


\begin{thebibliography}{}

\bibitem[]{}
Alpar M.~A., Shaham J., 1985, Nat, 316, 239
\bibitem[]{}
Alpar M.~A., Hasinger G., Shaham J., Yancopoulos S., 1992, A\&A, 257, 627
\bibitem[]{}
Andrews D., 1984, {\it EXOSAT} Express, 5, 31
\bibitem[]{}
Andrews D., Stella L., 1985, {\it EXOSAT} Express, 10, 35
\bibitem[]{}
Berger M., van der Klis M., 1994, A\&A, 292, 175
\bibitem[]{}
Berger M., van der Klis M., 1997, A\&A, submitted
\bibitem[]{}
Bildsten L., 1993, ApJ, 418, L21
\bibitem[]{}
Bildsten L., 1995, ApJ, 438, 852
\bibitem[]{}
Bulik T., M\'{e}sz\'{a}ros P., Woo J.~W., Nagase F., Makeshima K., 1992, ApJ, 
395, 564
\bibitem[]{}
Cowley A.~P., Crampton D., Hutchings J.~B., 1979, ApJ, 231, 539
\bibitem[]{}
Crampton D., Cowley A.~P., Hutchings J.~B., Kaat C., 1976, ApJ, 207, 907
\bibitem[]{}
Deeter J.~E., 1984, ApJ, 281,482
\bibitem[]{}
Dieters S., van der Klis M., 1997, MNRAS, submitted
\bibitem[]{}
Focke W.B., 1997, ApJ, 470, L127
\bibitem[]{}
Fortner B., Lamb F.~K., Miller G.~S., 1989, Nat, 342, 775
\bibitem[]{}
Ghosh P., Lamb F.~K., 1979, ApJ 234, 296
\bibitem[]{}
Ghosh P., Lamb F.~K., 1992, in van den Heuvel E.~P.~J., Rappaport S.~A.,
eds, Proc.\ NATO Conf.\ on X-ray Binaries and 
Recycled Pulsars. Kluwer, Dordrecht, p.~487
\bibitem[]{}
Hasinger G., 1987a, in Helfand D.~J., Huang J.-H., eds, 
IAU Symp.\ 125, The Origin and Evolution of Neutron Stars. 
Kluwer, Dordracht, p.~333
\bibitem[]{}
Hasinger G., 1987b, A\&A, 186, 153
\bibitem[]{}
Hasinger G., Priedhorsky W.C., Middleditch J., 1989, ApJ, 337, 843
\bibitem[]{}
Hasinger G., van der Klis M., 1989, A\&A, 225, 79
\bibitem[]{}
Hasinger G., van der Klis M., Ebisawa K., Dotani T., Mitsuda K., 1990, A\&A, 
235, 131
\bibitem[]{}
Hertz P., Vaughan B., Wood K.~S., Norris J.~P., Mitsuda K., Michelson K.~P.,
Dotani T., 1992, ApJ, 396, 201
\bibitem[]{}
Hoffman J.~A., Marshall H.~L., Lewin W.~H.~G., 1978, Nat, 271, 630
\bibitem[]{}
Jongert H., van der Klis M., 1996, A\&A, 310, 474
\bibitem[]{}
Kahn S.~M., Grindlay J.~E., 1984, ApJ, 281, 826
\bibitem[]{}
Kuulkers E. , 1995, PhD thesis, University of Amsterdam
\bibitem[]{}
Kuulkers E., van der Klis M., 1995, A\&A, 303, 801
\bibitem[]{}
Kuulkers E., van der Klis M., 1996, A\&A, 314, 567
\bibitem[]{}
Kuulkers E., van der Klis M., Oosterbroek T., Asai K., Dotani T., 
Van Paradijs J., Lewin W.~H.~G., 1994a, A\&A, 289, 795
\bibitem[]{}
Kuulkers E., van der Klis M., Oosterbroek T., 
Van Paradijs J., Lewin W.~H.~G., 1994b, in Holt S.~S., Day C.~S., eds,
The Evolution of X-ray Binaries. American Institute of Physics, p.~539
\bibitem[]{}
Kuulkers E., van der Klis M., Van Paradijs J., 1995, ApJ, 450, 748
\bibitem[]{}
Kuulkers E., van der Klis M., Vaughan B.~A., 1996, A\&A, 311, 197
\bibitem[]{}
Lamb F.~K., 1989, Hunt J., Battrick B., eds,
23rd ESLAB Symp.\ on Two-Topics in X-ray Astronomy: X-ray Binaries.
ESA SP-296, p.~215
\bibitem[]{}
Lamb F.~K., Shibazaki N., Shaham J., Alpar M.~A., 1985, Nat, 317, 681
\bibitem[]{}
Langmeier A., Hasinger G., Tr\"{u}mper J., 1990, A\&A, 228, 89
\bibitem[]{}
Langmeier A., Sztajno M., Vacca W.~D., Tr\"{u}mper J., Pietsch W., 1986, 
in Tr\"{u}mper J.\ et al., eds, The Evolution of Galactic X-ray Binaries.
D.~Reidel Publishing Company, p.~253
\bibitem[]{}
Leahy D.~A., 1991, MNRAS, 251, 203
\bibitem[]{}
Leahy D.~A., Darbro W., Elsner R., Weisskopf M.~C., Sutherland P.~G.,
Kahn S., Grindlay J.~E., 1983, ApJ, 266, 160
\bibitem[]{}
Lewin W.~H.~G., Lubin L.~M., Tan J., van der Klis M., Van Paradijs J.,
Penninx W., Dotani T., Mitsuda K., 1992, MNRAS, 256, 545
\bibitem[]{}
Lewin W.~H.~G., Van Paradijs J., Taam R.~E., 1993, Sp.\ Sc.\ Rev., 62, 223
\bibitem[]{}
Lewin W.~H.~G., Van Paradijs J., van der Klis M., 1988, Sp.\ Sc.\ Rev., 46, 273
\bibitem[]{}
Penninx W., Lewin W.~H.~G., Mitsuda K., van der Klis M., Van Paradijs J., 
Zijlstra A.~A., 1990, MNRAS, 243, 114
\bibitem[]{}
Penninx W., Lewin W.~H.~G., Tan J., Van Paradijs J., van der Klis M., 
Mitsuda K., 1991, MNRAS, 249, 113
\bibitem[]{}
Psaltis D., Lamb, F.K., Miller G.S., 1995, ApJ, 454, L137
\bibitem[]{}
Schoelkopf R.~J., Kelley R.~L., 1991, ApJ, 375, 696
\bibitem[]{}
Schulz N.~S., Hasinger G., Tr\"{u}mper J., 1989, A\&A, 225, 48
\bibitem[]{}
Smale A.~P., et al., 1996, in preparation
\bibitem[]{}
Spruit H.~C., Taam R.~E., 1990, A\&A, 229, 475
\bibitem[]{}
Stella L., Parmar A.~N., White N.~E., 1987, ApJ, 321, 418
\bibitem[]{}
Sztajno M., Van Paradijs J., Lewin W.~H.~G., Langmeier A., Tr\"{u}mper J., 
Pietsch W., 1986, MNRAS, 222, 499
\bibitem[]{}
Taam R.~E., Woosley S.~E., Lamb D.~Q., 1996, ApJ 459, 271
\bibitem[]{}
Tawara Y., Hirano T., Kii T., Matsuoka M., Murakami T., 1984, PASP, 36, 861
\bibitem[]{}
Tennant A.~F., 1987, MNRAS, 226, 963
\bibitem[]{}
Turner M.~J.~L., Smith A., Zimmerman H.~U., 1981, Sp.\ Sci.\ Rev., 30, 513
\bibitem[]{}
van der Klis M., 1989, in \"{O}gelman H., van den Heuvel E.~P.~J., eds,
Timing Neutron Stars. Kluwer, Dordrecht, p.~27
\bibitem[]{}
van der Klis M., 1994, ApJS, 92, 511
\bibitem[]{}
van der Klis M., Stella L., White N., Jansen F., Parmar A.~N., 1987, ApJ, 
316, 411
\bibitem[]{}
van der Klis M., Swank J.~H., Zhang W., Jahoda K., Morgan E.~H., Lewin W.~H.~G.,
Vaughan B., Van Paradijs J., 1996, ApJ, 469, L1
\bibitem[]{}
Van Paradijs J., Hasinger G., Lewin W.~H.~G., van der Klis M., Sztajno M., 
Schulz N., Jansen F., 1988, MNRAS, 231, 379
\bibitem[]{}
Van Paradijs J., Penninx W., Lewin W.~H.~G.,
Stzajno M., Tr\"{u}mper J., 1988, A\&A, 192, 147
\bibitem[]{}
Vaughan B.~A., van der Klis M., Wood K.~S., Norris J.~P., Hertz P., 
Michelson P.~F., Van Paradijs J., Lewin W.~H.~G., Mitsuda K., Penninx W., 
1994, ApJ, 435, 362
\bibitem[]{}
White N., 1986, {\it EXOSAT} Express, 16, 2
\bibitem[]{}
White N.~E., Peacock A., 1988, Mem.\ S.\ A.\ It., 59, 7
\bibitem[]{}
Wijnands R.~A.~D., Kuulkers E., Smale A.~P., 1996b, ApJ, 473, L45
\bibitem[]{}
Wijnands R.~A.~D., van der Klis M., Kuulkers E., Asai K., 
Hasinger G., 1997, A\&A, in press
\bibitem[]{}
Wijnands R.~A.~D., van der Klis M., Psaltis D., Lamb F.~K., 
Kuulkers E., Dieters S., Van Paradijs J., Lewin W.~H.~G., 1996a, ApJ, 469, L5

\end{thebibliography}
\end{document}